\input harvmac

\def\ZZ{\relax\ifmmode\mathchoice
{\hbox{\cmss Z\kern-.4em Z}}{\hbox{\cmss Z\kern-.4em Z}}
{\lower.9pt\hbox{\cmsss Z\kern-.4em Z}}
{\lower1.2pt\hbox{\cmsss Z\kern-.4em Z}}\else{\cmss Z\kern-.4em Z}\fi}

\def\a{\alpha}

\def\m\mu 
\def\n{\nu}
\def\s{\sigma}

\def\p{\partial}

\def\CZ{{\cal Z}}
\def\CS{{\cal S}}

\def\CA{{\cal A}}

\def\CG{{\cal G}}
\def\CD{{\cal D}}
\def\R{\relax{\rm I\kern-.18em R}}
\font\cmss=cmss10 \font\cmsss=cmss10 at 7pt

\def\hf{{1\over 2}}


\parindent 25pt
\overfullrule=0pt
\tolerance=10000

\sequentialequations


\def\lr{\lref}

\def\npb#1(#2)#3 {Nucl. Phys. {\bf B#1} (#2) #3 }
\def\rep#1(#2)#3 {Phys. Rept.{\bf #1} (#2) #3 }
\def\plb#1(#2)#3{Phys. Lett. {\bf #1B} (#2) #3}
\def\prl#1(#2)#3{Phys. Rev. Lett. {\bf #1} (#2) #3}
\def\physrev#1(#2)#3{Phys. Rev. {\bf D#1} (#2) #3}
\def\ap#1(#2)#3{Ann. Phys. {\bf #1} (#2) #3}
\def\rmp#1(#2)#3{Rev. Mod. Phys. {\bf #1} (#2) #3}
\def\cmp#1(#2)#3{Commun. Math. Phys. {\bf #1} (#2) #3}
\def\mpl#1(#2)#3{Mod. Phys. Lett. {\bf #1} (#2) #3}
\def\ijmp#1(#2)#3{Int. J. Mod. Phys. {\bf A#1} (#2) #3}
\def\jhep#1(#2)#3{JHEP {\bf #1} (#2) #3}
\def\jmp#1(#2)#3{J. Math. Phys. {\bf #1} (#2) #3}

\lr\rfBFSS{T.~Banks, W.~Fischler, S.~Shenker and L.~Susskind, 
\physrev55(1997)5112, hep-th/9610043.}
\lr\rfIKKT{N.~Ishibashi, H.~Kawai, Y.~Kitazawa and A.~Tsuchiya, 
\npb498(1997)467, hep-th/9612115.}
\lr\rfGGI{M.B.~Green and M.~Gutperle, \jhep01(1998)005, hep-th/9711107.}
\lr\rfDVV{R.~Dijkgraaf, E.~Verlinde  and H.~Verlinde, \npb500(1997)43,
  hep-th/9703030.}
\lr\rfKNS{V.~Krauth, H.~Nicolai and  M.~Staudacher, \plb431(1998)31,
  hep-th/9803117.}
\lr\rfMNS{G.~Moore, N.~Nekrasov and S.~Shatashvili, hep-th/9803265.}
\lr\rfYi{P.~Yi, \npb505(1997)307, hep-th/9704098}
\lr\rfSethiStern{S.~Sethi and M.~Stern, \cmp194(1998)675, hep-th/9705046.}
\lr\rfWittenbound{E.~Witten, \npb460(1996)335, hep-th/9510135.} 
\lr\rfGGIII{M.B.~Green and M.~Gutperle, \npb498(1997)195, hep-th/9701093.}
\lr\rfGreenVanhove{M.B.~Green and P.~Vanhove, \plb408(1997)122, 
hep-th/9704145; C.~Bachas, C.~Fabre, E.~Kiritsis, N.A.~Obers and 
P.~Vanhove, \npb509(1998)33, hep-th/9707126; B.~Pioline and E.~Kiritsis, 
\npb508(1997)509, hep-th/9707018; \plb418(1998)61, 
hep-th/9710078; E.~Kiritsis and N.A.~Obers, \jhep10(1997)004, 
hep-th/9709058; I.~Antoniadis, B.~Pioline and  T.R.~Taylor, 
\npb512(1998)61, hep-th/9707222.}
\lr\rfOoguriVafa{H.~Ooguri and C.~Vafa, \prl77(1996)3296, hep-th/9608079.}
\lr\rfBeckers{K.~Becker, M.~Becker and A.~Strominger, \npb456(1995)130,
  hep-th/9507158.}
\lr\rfBachaslecture{C.~Bachas, Nucl.Phys.Proc.Suppl. {\bf 68} (1998) 348, 
hep-th/9710102.}
\lr\rfLerche{W.~Lerche, \npb308(1988)102.}
\lr\rfMNVW{J.A.~Minahan, D.~Nemeschansky, C.~Vafa and N.P.~Warner, 
hep-th/9802168.} 
\lr\rfVanhoveCargese{P.~Vanhove, Carg{\`e}se May-June 1997, 
hep-th/9712079; {\sl Au bout de la corde... la th{\'e}orie M\/}, PhD
  dissertation, {\'E}cole polytechnique, April 1998, CPTH-T602/0498.}
\lr\rfPilch{K.~Pilch, A.N.~Schellekens and N.P.~Warner, \npb287(1987)362;
\ W. Lerche, B.E.W. Nilsson, A. N. Shellekens and N.P. Warner, \npb299(1988)91; \ 
 W. Lerche, \npb308(1988)102.}
\lr\rfRey{S.J.~Rey, \npb502(1997)170, hep-th/9704158.}
\lr\rfLowe{D.A.~Lowe, \plb403(1997)243, hep-th/9704041.}
\lr\rfBanksMotl{T.~Banks and L.~Motl, \jhep12(1997)004, hep-th/9703218.}
\lr\rfTseytlin{E.~Fradkin and A.~Tseytlin, \plb158(1985)316; {\sl ibid.\/}
  {\bf 160B} (1985) 69.}
\lref\rfMotl{L.~Motl, hep-th/9701025.}
\lref\rfBSB{T.~Banks and N.~Seiberg, \npb497(1997)41, hep-th/9702187.}
\lr\rfPartouche{A.~Kehagias and H.~Partouche, hep-th/9712164.}
  \lr\rfKS{ V.~Krauth and M.~Staudacher, hep-th/9804199.}
\lr\rfDhokerPhong{E.~D'Hoker and D.H.~Phong, \rmp60(1988)917.}
\lref\Witdgt{ E.~ Witten, \cmp141(1991)153.}
\lref\rfDMVV{R.~Dijkgraaf, G.~Moore, E.~Verlinde and H.~Verlinde, \cmp185(1997)197, hep-th/9608096.}
\lr\rfBachasLec{C.~Bachas, Nucl.Phys.Proc.Suppl. {\bf 68} (1998) 348, 
hep-th/9710102.}
\lr\rfGHV{S.~Giddings, F.~Hacqueboard and H.~Verlinde, hep-th/9804121.}
\lr\rfBonora{G.~Bonelli, L.~Bonora and F.~Nesti, hep-th/9807232.}
\lr\rfTom{T.~Wynter, \plb415(1997)349 , hep-th/9709029.}
\lr\rfTombis{T.~Wynter, hep-th/9806173.}
\lr\rfBCAP{M. Bill{\'o}, M. Caselle, A. D'Adda and P. Provero, hep-th/9809095}


\rightline{hep-th/9809130}
\rightline{CERN-TH/98-280}
\rightline{CPTH-S587.1297}
\rightline{DAMTP-1998-119}
\rightline{SPHT/98/097}
\Title{}
{\vbox{ \centerline {Matrix String Partition Functions}
\centerline{}}}
 
\centerline{Ivan K.~Kostov\footnote{$^\ast$}{Member of CNRS}\footnote{$ 
^\diamond $}{Permanent address:  C.E.A. - Saclay, Service de 
physique th{\'e}orique,  F-91191 Gif-sur-Yvette, France}${}^\dagger$
\ {\it and}  \ 
Pierre Vanhove\footnote{$^\bullet$}{Address after 1 October 1998:
  DAMTP, Cambridge University, Cambridge CB3 9EW, UK}\footnote{$ 
^\dagger$}{\tt ivan.kostov, pierre.vanhove@cern.ch}}

\centerline{{\it Theory Division, CERN}}
\centerline{{\it 1211 Geneva 23, Switzerland}}

\vskip .3in

\baselineskip8pt
{ 
We evaluate quasi-classically the Ramond partition function of Euclidean
$D=10$ $U(N)$ super-Yang--Mills theory reduced to a two-dimensional torus. The
result can be interpreted in terms of free strings wrapping the space-time
torus, as expected from the point of view of Matrix string theory. We
demonstrate that, when extrapolated to the ultraviolet limit (small area of
the torus), the quasi-classical expressions reproduce exactly the recently
obtained expression for the partition function of the completely reduced  SYM
theory, including the overall numerical factor. This is an evidence that our
quasi-classical calculation might be exact.  
 }
 
\vskip .3in
\vfill
\line{CERN-TH/98-280\hfill}
\Date{ September 1998 }

\baselineskip=16pt plus 2pt minus 2pt
\baselineskip=20pt plus 2pt minus 2pt

  
\newsec{Introduction}

The interpretation \refs{\rfWittenbound} of the dimensional reductions of
ten-dimensional supersymmetric Yang--Mills (SYM) theory as  effective theories
for the dynamics of $p$-dimensional extended objects (D$p$-branes) initiated a
new wave of interest in these theories. It culminated in the BFSS conjecture
\refs{\rfBFSS} that a system of interacting D0-branes, described by the
$D=10$ SYM theory reduced to one dimension, provides, in the large $N$ limit,
a constructive definition of M-theory, the hypothetical theory encompassing all
known string theories and eleven-dimensional supergravity. It also led to
Matrix string theory \refs{\rfMotl,\rfBSB,\rfDVV}, which  describes
non-perturbatively  type IIA string theory  by $D=10$ SYM theory reduced to
two dimensions. Finally, an interpretation of the completely reduced SYM
theory as type~IIB string theory has been advanced  in \refs{\rfIKKT}. All
three reduced SYM theories are closely related and, in the large $N$ limits
each one contains, in a certain sense,   the other two.  We will refer to the
SYM theory reduced to $2, 1$ and $ 0$ dimensions as the DVV, BFSS and  IKKT
model, correspondingly.  

The most basic information about these theories, namely  concerning their
vacuum excitations, can be obtained by studying their partition 
functions.\foot{By partition function of a supersymmetric matrix model we
understand the volume form in the sector  with  a minimal number of fermionic  zero-modes.}
 The only  partition function computed  at present is that of the completely reduced theory, which we will denote by $\CZ_{ \rm IKKT}$. It was studied by several groups \refs{\rfYi,\rfSethiStern,\rfKNS,\rfMNS} in order to prove the existence of bound states in the BFSS model \refs{\rfYi,\rfSethiStern,\rfGGI}. It was conjectured by   Green and Gutperle \refs{\rfGGI}  that 
\eqn\eResult{ 
{\CZ}_{_{\rm IKKT} } (g) =  g^{-{7\over 2}(N^2-1)}    \CF_N \sum_{m|N} {1\over m^{2}}, }
where the numerical factor $\CF_N$ was computed later  by Krauth, Nicolai and Staudacher \refs{\rfKNS}.
Recently, this  conjecture  was  rigorously proved by Moore, Nekrasov and Shatashvili \refs{\rfMNS}.

In  this paper we present  the quasi-classical calculation of the partition
function of the Euclidean matrix string theory (the DVV model) compactified on
a rectangular torus $\CT^2$ with periods $R$ and $T$, with  Ramond-Ramond
boundary conditions.\foot{The gauge coupling constant can be absorbed in the
  area of the torus and therefore is considered to be equal to 1.}
Replacing  the fermionic and bosonic potentials by constraints, we obtain 
\eqn\empF{\CZ_{_{\rm DVV} }=   \sum_{m|N}{1\over 
m}\ \sum_{p\in \ZZ} e^{-  { RT\over 2} N  p^2}.
}
\smallskip
Further we argue that this expression which, by its definition, is valid only
in the limit of large area $RT$, is actually exact.  
Our argument is based on the quasi-classical calculation of the DVV  partition
function in which  the constant mode $A_\s^{(0)}$ of the $U(1)$-component of
the gauge field subtracted by inserting a delta-function, and which we denote
by $\langle \delta(A_\s^{(0)}/\sqrt{2\pi})\rangle_{_{\rm DVV}}$.  In the
limit $RT\to 0$ the modified partition function coincides with the
partition function~\eResult\ of the completely reduced theory
\eqn\DVVIKKTa{\left\langle
\delta\left(A^{(0)}/\sqrt{2\pi g}\right)\right\rangle _{_{\rm DVV} }
\ \ 
\longrightarrow_{_{_{\!\!\!\! \! \! \! \!\!\!\!\!\!\!\!\!\! R, T\to 0}} }
 \ \ \   \ { (RT)^{-{7\over 2}(N^2-1)}
\over T\ \CF_N} \  \CZ_{_{\rm IKKT}}
  \left({1/ RT}\right) .}
  Comparing  our quasi-classical result 
\eqn\empfDa{
   \left\langle  \delta\left({A_\s^{(0)}/
      \sqrt{2\pi }}\right) \right\rangle_{_{\rm DVV} } = {1\over T}  \sum_{m|N} 
{1\over m^2}
\  \sum_{E\in \ZZ} e^{ -\hf{ E^2\over  RT N} } 
 .}
extrapolated to the $RT\to 0$ limit with the exact expression~\eResult\ we
find perfect agreement, including the numerical factor computed in
\refs{\rfKNS,\rfKS}.  It is therefore very plausible that our quasi-classical
results are valid everywhere, {\sl i.e.\/}, that these models possess the
property of having {\it exact quasi-classics}. 

The excitations that contribute to the DVV partition function can be
interpreted as free type IIA Green-Schwarz strings wrapping the torus, with
additional abelian gauge degrees of freedom on the world-sheet. The sum over
all possible wrappings  reproduces, in the large $N$ limit, the integration
over the string moduli.


The paper is organized as follows.  In Section~2, we define the DVV partition
function and make a quantitative description of the dimensional reductions
that give the BFSS and IKKT partition functions.  In Sections 3 and 4, we
present the derivation of~\empF\ and~\empfDa. Section~5 contains 
discussions of our results.   


\newsec{The  partition functions of the DVV, BFSS and IKKT matrix 
models}

\subsec{The functional integral of Matrix string theory (the DVV model)
compactified on a torus}

The Euclidean DVV model  is ten-dimensional $U(N)$ SYM theory dimensionally
reduced to a two-dimensional cylinder \refs{\rfDVV}.  The field content of the
theory includes  the two-dimensional $U(N)$ gauge field $\{ A_a\} _{a=1}^{2}$,
the bosonic Hermitian Higgs fields $\{X^I\}_{I=1}^{ 8}$, and the  Majorana--Weyl
fermions $\{\Psi_\alpha\}_{\alpha=1}^{16}$.  In order to study the partition
function, we compactify  the theory on a rectangular\foot{The matrix
  string action assumes that the metric on the two-dimensional space-time is
  $g_{ab}= \delta_{ab}$.} two-torus $\CT^2$ with periods $R$ and $T$, 

$$\CT^2  =\{ \vec \s = (\s, \tau) \ | \ \s \in [0, R], \tau \in [0,T]\}.$$
 The DVV partition  function is defined  by the functional integral 
\smallskip 
\eqn\eDefZ{
\CZ _{_{\rm DVV} } = \int { \CD A_a\over {\rm Vol} (\CG)}
\  \ \CD X  \CD \Psi \ \ 
 \prod_{I=1}^{8} \delta\left({X_I^{(0)}\over \sqrt{2\pi}}\right) 
    \prod_{\alpha =1}^{16}\Psi_\alpha^{(0)} \ \ 
\exp\big(-\CS_{_{\rm DVV}}[A,X,\Psi]\big) }
\smallskip
where the action is 
\eqn\esYM{\eqalign{
\CS _{_{\rm DVV}}&= \int_{\CT^2} d^2\sigma \Tr  \Big[{F_{ab}^2\over 4} 
+{1\over2}\left[D_a,
    X_I\right]^2+{i\over2} {}\Psi^T [ D\!\!\!\! /\  ,\Psi]\cr
    &+
  { 1\over 2}  {}\Psi^T \Gamma_I [X_I,\Psi]
 -{1\over 4} \sum_{I,J} [X_I,X_J]^2  \Big],\cr}
}
$D_a = \partial_a  - i A_a$  
is the covariant derivative, and  a minimal set of pairs of fermionic and
bosonic zero-modes: 
\eqn\efzerO{
\Psi_\alpha^{(0)}= {\Tr \int_{\CT^2} d^2\sigma  \Psi_\alpha \over \sqrt{ N RT
}}, \
\ \ \ X_I^{(0)}= {\Tr \int _{\CT^2} d^2\sigma  X_I \over \sqrt{ N RT}}. 
}
is subtracted from the  integration measure of the matter fields.  The
integration measure over the gauge fields is divided as usual  by  the volume
$ {\rm Vol} (\CG)$ of the gauge group.    
  
 
\subsec{ Relation to the partition function of the BFSS matrix model
  compactified on a circle} 

After shrinking one of the periods of the torus, the theory degenerates to a
SYM theory reduced to one dimension. In the limit $R\to 0$, the component
$A_\s$ of the gauge field  enters in the action in the same way as the eight
Higgs fields, and the $O(8)$ symmetry of the action is enhanced to $O(9)$.
The functional integral~\eDefZ\ describes, in this limit, the  partition
function of a one-dimensional reduction of the SYM theory, with $A_\s$ playing
the role of $X_9$.  However, this is not yet the partition function of the
BFSS model because the {\it integration measure\/} of the field $A_\s$   is not
identical to that of the eight remaining Higgs fields. In order to obtain the
same integration measure, we have to introduce in  the original functional
integral~\eDefZ\ a delta-function of the constant mode 
\eqn\xdevo{A_\s ^{(0)}=
  {\Tr \int_{\CT^2} d^2 \s  A_\sigma  \over \sqrt{N RT}}}
$\s$-component of the gauge field. Let us  denote  by $\langle \delta(A_\s
^{(0)}/\sqrt{2\pi})\rangle _{_{\rm DVV}} $ the DVV functional
integral~\eDefZ\ with the inserted delta-function. In the limit $R\to 0$ it
indeed reduces to  the partition function $\CZ_{_{\rm BFSS}}$ of the   BFSS
model at finite temperature $T$, defined by the action 
%
\eqn\Eibftc{
\CS _{_{\rm BFSS}} = R
\int_0^T \!\!\! d\tau\,  \Tr\Big({1\over2}\left[D_\tau ,X_I\right]^2  + 
{i\over2}\,{}\Psi^T
  [D_\tau,\Psi] -{1\over 4} \sum_{I,J} [X_I, X_J]^2  +{1\over 2}
\,{} \Psi ^T \Gamma^I[X_I, \Psi]\Big),
}
\smallskip
 \eqn\DVVBGSS{   \left\langle \delta\left({A_\s ^{(0)}/ \sqrt{2\pi }}\right)\right\rangle
  _{_{\rm DVV}}\ \ \  \rightarrow_{_{_{\!\!\!\!\!\!\!\!\!\!\! R\to 0}} }
 \ \ \  \CZ_{_{\rm BFSS}} .}
 \bigskip
 
  \subsec{The IKKT model as the high-temperature limit of the BFSS model}

The IKKT matrix integral  
\eqn\eIKKT{
\CZ_{_{\rm IKKT}} (g)  = \prod_{\mu =1}^{10}
[dX^\mu ] 
 \prod_{\alpha 
=1}^{16} [d\Psi_\a] 
 \ \exp\left(-  {1\over g}  \CS_{_{\rm IKKT}}[X,\Psi]\right)}
with the action
\eqn\eIKKTa{
\CS_{_{\rm IKKT}} = - {1\over 4}\sum_{\mu ,\nu} \Tr\,[X_\mu , X_\nu]^2
+ {1\over 2} \sum_\mu    \Tr\, {}\Psi^T[\Gamma^\mu X_\mu , 
\Psi] .
}
depends on a single parameter, the coupling constant $g= g_{\rm YM}^2$.  In refs. \refs{\rfKNS,\rfMNS} the integral was understood as an integral  over traceless matrices.  Here we   use the same normalizations as in \refs{\rfKNS} but, in
order to facilitate  the comparison with the  previous integrals,  we define the integration measures by 
\eqn\eFLFL{ [dX^\mu ]  \ =\ dX^\mu \ 
\delta \Big( {\Tr X^\mu \over \sqrt{ 2\pi N}}\Big) , \ \ \ \ 
 [d\Psi_\a]  \ =\ d\Psi_\a \ {\Tr \Psi_\a  \over \sqrt{N}},
}
where  $dX $ and $d\Psi$ are the   flat measures with normalization 
\eqn\IKKTnorm{
 \int d X e^{-{1\over 2} \Tr\,  X^2} =1, \ \ \ \
\int d \Psi_\alpha d \Psi_\beta  e^{-\Tr\,{}\Psi_\alpha^T \Psi_\beta }=1.
}
The partition function of the IKKT model, with the above normalization of 
the measure, is equal to \refs{\rfKNS,\rfMNS}\foot{The overall power of $g$
can be determined by dimensional arguments. The integration over the 
fermions gives a pfaffian, which is a homogeneous polynomial 
in $X/g$ of degree $8(N^2-1)$. The rescaling $X\to g^{1/4}X$   makes the action $g$-independent and produces a factor  $g^{(N^2-1)( 8(1- 1/4)-10/4)}=g^{7/2(N^2-1)}$.}
\eqn\eKNS{\CZ_{_{\rm IKKT}}(g)\ =\   g^{-{7\over 2}(N^2-1)}  \  \CF_N  \  
\sum_{m|N} {1\over m^2}}
where $\CF_N$ is the   ``group factor" \rfKNS 
\eqn\eFN{\CF_N=   { \sqrt{N}\ 
N! \over (\sqrt{\pi})^{N^2-1}}\ \ 
  {1\over 2\pi N}  \prod_{k=1}^N {(2\pi)^k\over k!}
 .}

The $T\to 0$ limit of the BFSS model has been  analysed by Sethi and Stern
\refs{\rfSethiStern,\rfGGI} in connection with the computation of the Witten
index. The BFSS action \Eibftc\ reduces,  in the limit $T\to 0$ and after
identifying the  constant mode of the gauge field    with  the Higgs field
$X_{10}$, to the $O(10)$-symmetric IKKT action~\eIKKTa\  multiplied by the
area $RT$ of the torus. The limit for the measure is less trivial. Let us
first consider the measure of the gauge field. 

The one-dimensional  gauge field  $A$ corresponds, by the exponential map,
to a generic element of the  local  gauge group $U (\tau)= \hat T
e^{i\int_0^\tau  A(t) dt}$. It is therefore convenient to parametrize the
constant mode $A$ of the gauge field in terms of the group element $U =
e^{iTA} $ and integrate over the Haar measure on $SU(N)$ (normalized as
$\int_{SU(N)} dU =1$).  In the vicinity of any of the $N$ central elements
of $SU(N)$ the group element can be parametrized by  $U= e^{2\pi i k/N}
e^{iTA} $ and the Haar measure becomes 
\eqn\eGA{ {dU\over {\rm Vol} [SU(N) ] } \ \to  \  { T^{N^2-1}\over N\CF_N}  [dA   ]
}
where $\CF_N$ is given by~\eFN \ and the measure $[dA]$ is
identical to the measure $[dX]$  defined in~\eFLFL . (The details of the
calculation can be found in \refs{\rfKS}.)
As was explained by Sethi and Stern \refs{\rfSethiStern}, in the limit $T\to 0$ the integral  over the gauge   field  is saturated by the  vicinity of the $N$ central elements  of $SU(N)$ and  therefore by eq.~\eGA\ 
\eqn\eMeasA{ \CD A \ \ \ 
\rightarrow_{_{_{\!\!\!\!\!\!\!\!\!\!\! T\to 0}} }
 \ \ \ \  { T^{N^2-1}\over \CF_N} [dX_{10}]. }
For the measures of the matter fields we find,
from the kinetic part of the action \Eibftc\ 
and the normalization \IKKTnorm ,
\eqn\eMeasurE{\eqalign{
\CD X &\to  \left({T\over R}\right) ^{-{9\over 2}N^2}\prod_{I=1}^9  dX_I\cr
\CD\Psi  & \to R^{- {16\over 2}N^2} \prod_{\a} d\Psi_\a .\cr}}
 Taking into account the rescaling of the zero-modes
\eqn\eMeaserE{\eqalign{
\prod _I \delta  \left({ X^{(0)}_I\over \sqrt{2\pi}}\right)
 &\to (RT)^{-{9\over 2}} \prod_I \delta \left(
{\Tr X_I\over \sqrt{2\pi N}}\right)\cr
\prod _\a \Psi ^{(0)}_\a  \ \ \ \ 
 &\to (RT)^{{16\over 2}} \ \prod_\a \delta \left(
{\Tr \Psi_\a\over \sqrt{  N}}\right)\cr}}
 and combining all   factors of $T$ and $R$, we 
finally obtain
\smallskip
  \eqn\DVVIKKT{\eqalign{ \left\langle
\delta\left(A^{(0)}/\sqrt{2\pi g}\right)\right\rangle _{_{\rm DVV} }
\ \ 
\longrightarrow_{_{_{\!\!\!\! \! \! \! \!\!\!\!\!\!\!\!\!\! R, T\to 0}} }
 \ \ \  & \ { (RT)^{-{7\over 2}(N^2-1)}
\over T\ \CF_N} \  \CZ_{_{\rm IKKT}}
  \left({1/ RT}\right) \cr
= 
  \ \ \ \ & \ {1\over T} \sum_{m|N}{1\over m^2}
   .\cr}} 
   \bigskip 
\bigskip 

%


\newsec{Quasi-classical calculation of the partition function of the DVV 
model}
 
 As argued by 
Dijkgraaf, Verlinde and Verlinde \refs{\rfDVV}, in 
 the infrared limit $ RT\to\infty$ the non-diagonal components of  the gauge
 and matter fields become infinitely massive, and  the 
bosonic and fermionic potentials turn into   
constraints.  Under the constraint that all matrices  $\Phi = \{ X_I, \Psi_\a, iD_a= i\p_a +A_a\} $
are simultaneously diagonalizable, for each field configuration
there exists a unitary matrix $V(\s,\tau)$ such that
 $$\Phi(\s, \tau)  =  V^{-1} (\s, \tau ) \Phi^D(\s, \tau )V(\s, \tau ),$$
where $\Phi^D = {\rm diag}\{\Phi_1,...,\Phi_N\}$.
We have therefore
\eqn\eBCpsn{\eqalign{
\Phi^D (R, \tau )&= \hat S ^{-1} \Phi^D ( 0, \tau )\hat  S  ,\cr
   \Phi^D ( \s, T  )&= \hat  T^{-1}   \Phi^D (\s, 0)\hat   T  
\cr}}
where $\hat S= V(0, \tau) V^{-1}(R, \tau )$ and $\hat  T=  V(\s, 0)V^{-1} (\s,
T )$. By construction, $\hat S\hat T = \hat T \hat S $. Assuming that all the
eigenvalues are distinct, the only unitary transformations relating two
diagonal matrices represent permutations of their diagonal elements. Therefore
the matrices $\hat S$ and $\hat T$ act as two commuting permutations $\hat s:
i\to s_i $ and $ \hat t: i \to t_i $ of the symmetric group $S_N$\ \foot{After
  our manuscript was finished, we learned about the paper \refs{\rfBCAP},
 where a similar
  interpretation of the partition function of {\it pure}  $U(N)$ YM theory on
  the torus is presented.}  

\eqn\eSTTS{\left[ \hat T ^{-1} \Phi \hat T\right]_i = \Phi _{t_i}, \ \ 
\left[ \hat  S ^{-1} \Phi \hat  S \right]_i = \Phi _{s_i}}
and, in particular, do not depend on the coordinates $\s$ and $\tau$.

Each pair of permutations describes an $N$-covering of the target-space torus
consisting, in general, of several connected components. Each connected
component can be interpreted as the world-sheet of a string  wrapping several
times the torus in both directions. A $q$-component covering corresponds to a
decomposition 
$$\hat s = \prod_{k=1}^q \hat s^{(k)},\ \ \hat t = \prod_{k=1}^q \hat
t^{(k)},$$
where all factors commute with each other and satisfy 
 \eqn\eBigtor{
(\hat s^{(k)})^{j_k} (\hat t^{(k)})^{m_k}= (\hat s^{(k)})^{n_k} (\hat
t^{(k)})^{l_k}=1, \ \ \ \sum_{k=1}^q (n_km_k - l_kj_k)  = N .}
The $k$-th world-sheet torus is the complex plane $\omega =\s + i\tau $
factored by the periods $\omega_1 = n_k R +i \ (l_k T) $ and $ \omega_2 =
j_k R +i\ (m_k T)$; it covers the target torus $N_k = n_km_k - l_kj_k$
times. In this way the orbifold structure of the target space  generates
the sum over all twisted boundary conditions, which becomes, in the limit
$N\to\infty$  and for the ``long" strings only, the integral over all
complex structure of the world-sheet tori.  
 Owing to the periodicity condition~\eBCpsn\ on the fermions, each connected
component has 16 fermionic zero modes. In our problem we are only
interested in contributions with exactly 16 fermionic zero-modes;
therefore only one-sheet $(q=1)$ coverings will be relevant. A covering
with periods $\omega_1 = n R +i (l T) $ and  $\omega_2 =j R+i (m T)$
wrapping the target torus $N = n m -jl$ times is defined by two
permutations, $\hat s$ and $\hat t$, satisfying
$$\hat s \hat
t=\hat t \hat s, \ \ \hat s^n \hat t^l=\hat s^j \hat t^m=1.$$
By a mapping class transformation we can reduce this to
\eqn\epeR{\hat s \hat t=\hat t
\hat s, \ \ \hat s^n =\hat s^j \hat t^m=1 \ \ \ (mn=N, j=0,1,...,n-1).}
The explicit solution of \epeR\ is, up to an internal homomorphism,
\eqn\eBCj{ \hat s=\{ i\to i + m ({\rm mod} \ N)\} , \ \ \hat t= \cases{
\{i \to i+ 1 ({\rm mod}\ m)\} & if $j=0$\cr
 \{i \to i- j ({\rm mod}\ N)\} & if $j=1, ..., n-1.$\cr}}
After having identified the distinct topological sectors, we can write the
partition function as a sum over all pairs $m$ and $n$ such that $mn=N$
and $j=0,1,..., n-1$.  The diagonal matrix variables
 $\Phi^D(\s,\tau)$, with boundary conditions belonging to the
 equivalence classes $[m, n; j]$, can be considered as scalar variables
defined on the torus with periods $\omega_1= nR$ and $\omega_2 
= jR +i (mT)$ in the $\omega = \s + i \tau$ plane.
  
Now let us proceed to the evaluation of the  partition function in the
infrared limit. In the limit $RT\to\infty$ the measure over the gauge field
reduces to the integral over the diagonal components of the gauge field,
normalized by the volume of the diagonal gauge group $ \CG^D$ which is the
localized $U(1)^{N}$. Taking into account the overall factor $1/N!$ because
the eigenvalues are determined up to a permutation, we get 
\eqn\eGmeas{
  \int{\CD A \over {\rm vol}(\CG)}\CD X \CD \Psi \rightarrow {1\over N!}
 \sum_{\hat s \hat t = \hat t \hat s}
\int  {\CD A^D \over {\rm vol}(\CG^D)}\CD X^D \CD \Psi^D}
where the boundary conditions in each term are twisted according to
eq.~\eSTTS. The number of permutations  in each class $[m,n;j]$  is equal  to
the number of combinations of $m$ and $n$ elements times the number of  cyclic
permutations of order $m$ and $n$,
$$  {N!\over n! m!}  (m-1)! (n-1)!  = (N-1)! .$$
   Summing the contributions of all equivalence classes we get 
\eqn\eZdvv{
\CZ_{_{\rm DVV}}  = { (N-1)!\over N!} \sum_{mn=N}\sum _{j=0, ...,
  n-1}\, \CZ_{[m, n; j]}
 }
where $\CZ_{[m, n; j]}$ is the partition function 
of the Abelian $(N=1)$ $DVV$ model defined on the 
  torus of area $NRT $ with periods $ \omega_1 =nR$
and $\omega_2 =jR + i( mT)$.  
The latter 
is a product of the partition function of the Abelian gauge field 
and that associated with the diagonal components of the matter field   %
\eqn\zZzZz{\CZ_{[m,n;j]} = \CZ_{[m,n;j]}^{^{\rm gauge}} \CZ_{[m,n;j]}^{^{\rm matter}}. 
}

 The contribution of the $U(1)$ gauge field  to the partition function  is given, in the gauge $A^D_\tau =0$, 
 to the functional integral  with respect to the    angular variable  
\eqn\thEta{\eqalign{\theta(\tau)
&= \int _0^{R} \Tr A_\s ^D(\tau, \s) d\s\cr
&=\int _0^{nR} A_\s (\tau, \s) d\s .\cr}}
Here $A_\s$ denotes the Abelian gauge field on the ``world-sheet" torus that
 corresponds to the $N$-component field  of $A^D_\s$ on the space-time torus. 

The gauge-field partition function is
\eqn\empfG{ \CZ_{[m,n;j]}^{\rm gauge} =
 \int_{ \!_{\theta (mT) = \theta (0)}}\!\!\!\!\!\!\!\!  \!\!\!\!\!\!\! \CD \theta
 \ e^{-{1\over 2 nR}
\int_0^{mT} d\tau  (\p_\tau \theta)^2  }  =  \sum _{p\in \ZZ} e^{- { 
RT\over 2} Np^2} . 
 }
The result of the integration depends only on the area $RT$ and  not on the
modular parameter of the torus, which reflects the 
symmetry of the two-dimensional gauge theory 
with respect to area-preserving diffeomorphisms. Further, with our conventions
for the zero modes, the integral of the matter fields
 is exactly 1 because of supersymmetry \refs{\rfDhokerPhong}

\eqn\edno{\CZ_{[m,n;j]}^{\rm matter}=1.}
This gives the result  \empF\  which can be written, after a Poisson resummation, as 
\eqn\empF{ \CZ_{_{\rm DVV}}   =\sum_{m|N}{1\over m}\ 
\sqrt{2\pi  \over  RT N}\sum_{E\in \ZZ} 
e^{-\hf {(2\pi E)^2 \over  RT N}}
. }

 
\newsec{The   DVV  partition function with subtracted constant mode
of the gauge field}

The evaluation of  the  modified DVV   \xdevo\ can be done in the same  way as
in the previous section. In the  topological sector $[m,n;j]$ the constant
mode $A_\s^{(0)}$  is expressed through the angular variable~\thEta\  as  
\eqn\aThta{A_\s^{(0)} = { \int _0^{mT}d\tau \theta (\tau)
\over\sqrt{ N RT}}.}
The combinatorics  is the same as in the previous section  and the only
difference is in the expression of the gauge-field partition function in each
topological sector. The latter is given  by the one-dimensional functional
integral  with respect to the field $\theta(\tau)$ taking its values in the
unit circle  
\eqn\empfGD{
 \tilde  \CZ_{[m,n;j]}=  
\int_{ \!_{\theta (mT) = \theta (0)}}\!\!\!\!\!\!\!\!  \!\!\!\!\!\!\! \CD \theta
 \ e^{-{1\over 2 nR}
\int_0^{mT} d\tau  (\p_\tau \theta)^2  } \ 
\delta\Big( {\int_0^{mT} d\tau \theta(\tau)
\over \sqrt{ 2\pi RT N  }}\Big),}
and  is evaluated as
\eqn\DERNIERE{  \tilde  \CZ_{[m,n;j]}= \sqrt{N RT\over 2\pi} {1\over mT} \sum_{p\in \ZZ} e^{-\hf RT N p^2}=
{1 \over mT}\ \sum_{E\in \ZZ} 
e^{-\hf{(2\pi E)^2\over RT N}}
.}

Summing over all topological sectors we find, instead of~\empF, 
\eqn\empfD{
   \left\langle  \delta\left({A_\s^{(0)}/
      \sqrt{2\pi }}\right) \right\rangle_{\rm DVV}  = {1\over T}  \sum_{m|N} 
{1\over m^2}
\  \sum_{E\in \ZZ} e^{ -\hf{ E^2\over  RT N} } 
 .}
Comparing this expression, whose 
derivation  implies  the infrared  limit $RT\to \infty$,
with Eq. \DVVIKKT,  we see that it is equally true 
in the ultraviolet limit $RT\to 0$.
 
 
\newsec{Discussion}
  
We have computed the quasi-classical partition function of the Euclidean Matrix
string theory compactified on a two-dimensional torus  $\CT^2$ , with doubly
periodic boundary conditions and a minimal set of zero-modes removed.  We have
shown that the  relevant degrees of freedom are described, in the limit of
large area of the torus, by an Abelian  supersymmetric sigma-model accompanied
by  a $U(1)$ gauge theory defined on the orbifold space $S^N \CT^2 =
(\CT^2)^{\otimes N}/S_N$, where  $S_N$ is the symmetric group of $N$
elements. The sigma-model and the gauge field are coupled through the boundary
conditions, which are described by a pair of commuting permutations of $S_N$.

Each such configuration can be interpreted as a set of non-interacting strings
in a light-cone gauge, with additional gauge degrees of freedom on the
world-sheet. The world-sheet of each of such string defines a multiple
covering of the space-time torus $\CT^2$,  characterized by a modular
parameter $\omega_2/\omega_1$, with $\omega_{1,2} $ sweeping the lattice $R\ZZ
+i T\ZZ$. In the large-$N$ limit, the sum over coverings converges to an
integral with a correct modular-invariant  measure, under the condition  that
we are far from the boundary of the moduli space.  This means that the
partition function of a ``long" string in the DVV model is indeed that of the
Green-Schwarz string in light-cone gauge, with additional Abelian gauge degree
of freedom on the world-sheet. To complete the proof of the (perturbative)
matrix-string correspondence, one has to consider the matrix-field
configurations that define branched coverings of the space-time torus which,
according to the original suggestion  in \refs{\rfDVV}, should describe
strings in interaction. Some steps in this direction was made by the authors
\refs{\rfGHV,\rfTom,\rfBonora}.   
 
On the other hand, near the boundary of the moduli space, that is  when one of
the periods is kept finite  while $N\to\infty$, the sum keeps its discrete
character and  the corresponding excitations (``short" strings) describe
particles. Thus the distinction between particle and string excitations in
the DVV model can be made only after performing the large-$N$ limit. The
latter is not yet completely understood, in spite of the recent progress made
in \refs{\rfTombis}. With the  convention that a minimal set of  zero-modes is
deleted from the functional measure,  which is the case  considered in the
present paper, the allowed field configurations of the orbifold theory are
described by a single string  whose world-sheet wraps $N$ times the
space-time torus $\CT^2$.  Our computation gives an intuitive understanding
of the  sum over the divisors of $N$ in the expression for the partition
function of the IKKT model~\eResult. It is quite analogous to the
argument presented by  Green and Gutperle in which this last partition function
was compared with the  partition function of the matrix quantum mechanics at
infinite temperature.  Their calculation   was based on the assumption of the
existence of  bound states of D0-branes, which lead to the sum over the
divisors of $N$. In our case the sum over the divisors comes from the sum
over the conjugacy classes of permutations defining the different topological
sectors of the $S_N$-orbifold theory. This is  a trivial example of the
Hecke operator, defined by its action on a modular-invariant form $\CA$
(actually a constant in our case)
 \eqn\hecke{
\CH_N [\CA]( \tau )=
 {1\over N} \sum_{N=mn\atop 0\leq j <n} \CA\left(m \tau +j\over n \right)
\  , }
which takes care of the inequivalent $N$-fold wrappings. (For more general
discussion see \refs{\rfDMVV,\rfBachasLec}.) Comparing our calculation with
the argument of Green and Gutperle, we see that the bound states of D0-branes
can indeed be interpreted as strings winding several times around the spatial
($\s$-)dimension of the two-torus.  

The partition functions of the DVV and the IKKT models are related to the
half-BPS saturated amplitudes, such as the $t_8t_8\CR^4$ term in type~IIB
theory \refs{\rfGGIII,\rfGreenVanhove}.  Our computation confirms the rules
for counting wrapped  D-branes used in
\refs{\rfGGIII,\rfGreenVanhove,\rfBeckers,\rfOoguriVafa}. Moreover, we have
shown that the high-temperature limit of the DVV model reproduces the
D-instantons contributions predicted in~\refs{\rfGGIII}, which  confirms the
dual interpretation of the DVV model as describing wrapped D1-branes. 
 
We consider as the principal result of this paper not the computation of the
IR partition function itself, which is rather  trivial, but its comparison
with the known result for the partition function of the  completely reduced
gauge theory (the IKKT model). For this purpose we have established the exact
relations between the partition functions of the SYM theories reduced to 2,
1 and 0 dimensions, which we presented in Section 2.  Then we were able to
check that  when extrapolated to the small-area limit, the quasi-classical
result reproduces exactly the known expression of the IKKT partition
function. Therefore we conjecture that expressions~\empF\ and~\empfD\ are
actually {\it exact\/}, which means, that the theory has the property of having
exact quasi-classics. We expect that this can be proved by extending to the
two-dimensional case the calculation  of  Moore, Nekrasov and
Shatashvili~\rfMNS\ based on Witten's description of the two-dimensional
gauge theory as cohomological field theory \refs{\Witdgt}.  

Finally, let us remark that  it is possible to give a very simple  explanation
of the pre-exponential factors in expression~\empF\ if the partition
function of the DVV model is  considered as a certain limit of the partition
function of a supersymmetric Schild string defined on the orbifold
$S^N\CT^2$. We intend to report on this subject in the near future.  

\bigbreak\bigskip\bigskip\centerline{{\bf Acknowledgements}}\nobreak
 
We thank Costas Bachas, Volodya Kazakov, Matthias Staudacher and  Tom Wynter
for many useful discussions. Partial support was received by I.K. from the
European contract TMR ERBFMRXCT960012. P.V. was supported by a short-term
visiting fellowship from CERN.  

\listrefs

\bye